# Aging and configurational memories at rough frictional interfaces

**Authors:** Kasra Farain[1], Daniel Bonn[1]

[1]Van der Waals–Zeeman Institute, Institute of Physics, University of Amsterdam; Science Park 904, 1098 XH Amsterdam, Netherlands.

## Abstract:

Frictional interfaces retain distinct memories of how they were sheared and how long they have rested, yet rate-and-state descriptions typically compress these histories into a single state variable. Here we combine nanometer-per-second sliding experiments on rough multi-asperity contacts with different protocols to disentangle temporal and configurational memories at frictional onset. We identify a broad friction force peak associated with shear-driven reorganization of the load-bearing contact network. By contrast, holding a previously sheared system at rest in an arrested configuration produces a much narrower friction force peak that grows approximately logarithmically with hold time and relaxes over nanometer-scale slip distances. Configurational disorder without waiting regenerates only the broad peak, whereas waiting without configurational disorder generates only the aging peak. These results show that aging and configurational disorder write and read experimentally separable frictional memories, and that static friction is a protocol-dependent quantity rather than a fixed material property.

## Introduction:

Following the classical work of Amontons and Coulomb, friction is usually presented as a simple material property: solids resist sliding up to a threshold set by a static friction coefficient, beyond which they slide with a lower kinetic friction coefficient (1, 2). In reality, frictional interfaces behave more like memory devices than timeless materials (3, 4). Nominally identical contacts can show very different resistance depending on how they were loaded in the past and how long they have remained at rest (5−10). This history dependence underpins stick–slip dynamics, the stability of engineered joints and bearings, and the friction laws used to model earthquake nucleation on geological faults (1, 2, 11−15).

A large body of work has shown that frictional contacts strengthen over time when held stationary, a process known as contact aging. This is most commonly observed in slide–hold–slide experiments, where the friction peak measured on restart grows approximately logarithmically with hold time (2, 5, 6, 8, 11, 12). This strengthening is attributed either to viscoelastic creep of load-bearing asperities, which can increase the real contact area (2, 5, 6), or to the gradual formation and strengthening of interfacial bonds at asperity contacts (16, 17). Both mechanisms act on junctions that remain stationary during the hold.

Sliding, by contrast, continuously deforms, breaks, and renews asperity microcontacts. It reorganizes the force-bearing contact network, an evolution set by accumulated slip and shear direction (5, 7, 18). In the contact-age interpretation of the Dieterich law, the time- and slip-dependent effects are linked by assuming that sliding sets the clock for aging (2, 5, 11, 12, 19, 20). Under shear, microcontacts are renewed after a characteristic slip distance, so their mean lifetime scales inversely with sliding speed. A more slowly sliding interface therefore carries, on average, older and correspondingly stronger junctions. In this picture, time and slip act on the same interfacial state. Standard rate-and-state formulations correspondingly encode hold-time strengthening and slip-driven renewal in the same internal state variable (11–14, 21, 22).

However, a one-state description cannot by itself establish whether time and slip act on the same microscopic degree of freedom or on distinct physical processes. Does sliding merely renew the age distribution of microcontacts, or does it also write an independent configurational memory of how forces are distributed across the contact network? Here we answer this question using nanometer-per-second sliding experiments in a vibration-isolated apparatus, combined with controlled slide–hold–slide, shear-arrest, contact-reset, and mechanical-perturbation protocols. We show that rough multi-asperity interfaces carry at least two coexisting and experimentally separable memories—an aging memory and a configurational memory. Their distinct transients at sliding onset reveal the limits of assigning a single microscopic interpretation to the frictional state variable and motivate descriptions that distinguish temporal and configurational memories.

## Results

### Configurational origin of the friction overshoot

Rough frictional interfaces transmit load through a disordered population of microscopic asperity contacts (1, 2, 5, 23). Resolving how this population evolves during sliding onset requires measuring friction and displacement with high resolution over slip distances long enough for the interface to approach steady state. We use a rheometer capable of applying rotational rates down to $10^{-8}$ $s^{-1}$. A custom rotary-to-linear fixture converts this rotation into translational sliding at speeds as low as 1 nm $s^{-1}$ (Methods; Fig. S1).

We previously reported a broad, nonmonotonic friction transient at sliding onset in this velocity range (7). Figure 1A shows the full onset transient for a polytetrafluoroethylene (PTFE)–glass interface driven at 2 nm $s^{-1}$. The friction coefficient initially increases with displacement (inset), resembling the slow-sliding strengthening transients described by rate-and-state friction (5, 14). With continued slip, however, friction decreases, producing a broad overshoot before reaching steady state. Similar overshoots are observed for rough steel–glass interfaces (Fig. 1B) and polypropylene–glass interfaces (Fig. S2). Previously published measurements on acrylic surfaces also contain weakly nonmonotonic responses resembling incomplete segments of the present overshoot, although the full peak was not resolved and the data were interpreted within a monotonic rate-and-state framework [Fig. 8 of ref. (5)]. The overshoot maximum is the largest shear resistance the interface sustains before reaching steady sliding; for the same interfacial preparation under force control, it sets the threshold for sustained macroscopic sliding, i.e., the static-friction threshold.

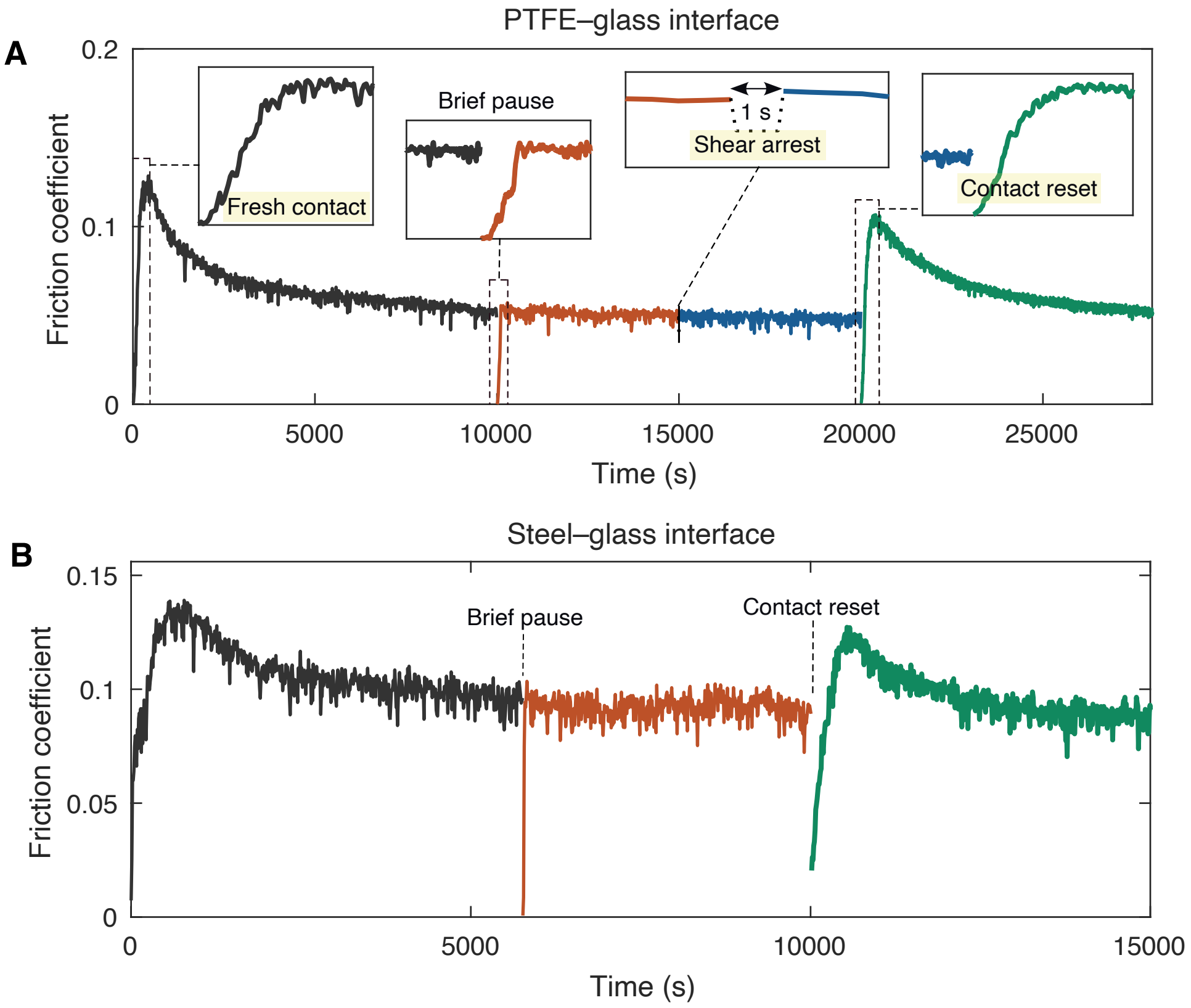

**Fig. 1. Friction overshoot at sliding onset and its configurational dependence.** (A) Friction coefficient versus time for a PTFE–glass interface driven at an imposed sliding velocity of 2 nm $s^{-1}$. Friction first rises, passes through a broad overshoot, and relaxes to the steady-sliding value $\mu_k$. The overshoot maximum is the largest resistance encountered before steady sliding and, under equivalent quasistatic force-controlled loading, corresponds to the static-friction threshold $\mu_s$. The left inset resolves the initial strengthening over 0–500 s, which resembles the slow-sliding strengthening transients described by rate-and-state friction. At $t$ = 10,000 s (second inset), sliding is paused for 5 s and then resumed; at $t$ = 15,000 s (third inset), velocity control is replaced for 1 s by force control at a shear force slightly below $F_{SS}$, arresting sliding; and at $t$ = 20,000 s (right inset), sliding is stopped and the surfaces are separated and recontacted before sliding resumes. Neither the brief pause nor the force-controlled arrest restores the broad overshoot, whereas separation and recontact does, consistent with the shear-prepared interfacial configuration surviving arrest but being reset by separation. (B) Corresponding experiment for a rough steel–glass interface driven at 4 nm $s^{-1}$. As in (A), a brief pause does not restore the broad overshoot, whereas separation and recontact does, demonstrating the same configurational dependence in a second material system.

What drives the friction overshoot? We consider the following physical picture. Macroscopic friction arises from the collective contribution of microscopic forces transmitted through surface asperities (24). It equals the vector sum of their in-plane components (Fig. S3). When two rough surfaces first come into contact, their asperities are disordered within the frictional interface and have no preferred orientation in the shear plane. In this state, the microscopic force contributions on average cancel in the shear plane. As shear is applied, asperities deform and move relative to one another, gradually biasing their force contributions along the sliding direction. Once steady sliding is reached, the sum of the in-plane components of the microscopic asperity forces equals the macroscopic kinetic friction force. We hypothesize that the friction overshoot emerges from this configurational transition of the asperity ensemble.

To test this hypothesis, we paused and resumed sliding in the steady-state sliding regime, using three protocols (Fig. 1A). In the first pause ($t$ = 10,000 s), we removed the applied sliding velocity and restarted it 5 s later. Sliding recommenced without a friction peak, suggesting that during the brief pause, asperities retained the configuration established by prior sliding rather than reverting to the disordered arrangement present before sliding began.

In the second pause ($t$ = 15,000 s), we switched from velocity control to force control, applying a shear force slightly below the steady-state (kinetic) friction $F_{SS}$. This halted sliding while preserving the stress state on the interface and keeping the asperity ensemble close to its steady-sliding configuration. When we switched back to velocity control 1 s later, no friction peak emerged. This indicates that the static-to-kinetic friction peak is absent when the interface is restarted from a configuration close to the steady-sliding state.

In the third pause ($t$ = 20,000 s), we separated the surfaces and brought them back into contact. This reset the interfacial configuration and restored the broad friction peak. Similar behavior was observed at rough steel–glass (Fig. 1B) and polypropylene–glass interfaces (Fig. S2). These results show that the overshoot appears when the initiation of sliding requires configurational reorganization of the asperity ensemble.

Finally, to change the interfacial configuration without interrupting sliding, we applied mechanical perturbations to the apparatus during steady-state sliding. A 200-g sand-filled pouch was dropped from approximately 20 cm onto the experimental table, transmitting a brief vibrational impulse through the apparatus without direct contact with the frictional interface or any change in the imposed rheometer loading protocol. Figure 2A shows the friction transient observed at an imposed sliding velocity of 1 μm $s^{-1}$, together with its regeneration following the perturbation. This transient resembles the weakening response conventionally observed at high sliding rates in rate-and-state friction (12, 14). At the much lower sliding velocity of 4 nm $s^{-1}$, the strengthening segment of the onset overshoot is also resolved, and applying the perturbation near the peak of this transient likewise regenerates the response (Fig. 2B). Thus, both the weakening and strengthening onset transients can be regenerated during continued sliding following a perturbation pulse, without a stationary hold. In contact-age interpretations of rate-and-state friction, these transients are associated with evolution of the same state variable that governs strengthening during stationary contact. Their regeneration during uninterrupted sliding therefore shows that a stationary aging period is not required for their recovery. The sensitivity to perturbations may also help explain why this configurational memory has been difficult to observe experimentally. Our apparatus is mounted on an optical vibration-isolation table, whereas ambient vibrations in many laboratory settings may be sufficient to disrupt the interfacial configuration during pauses in sliding.

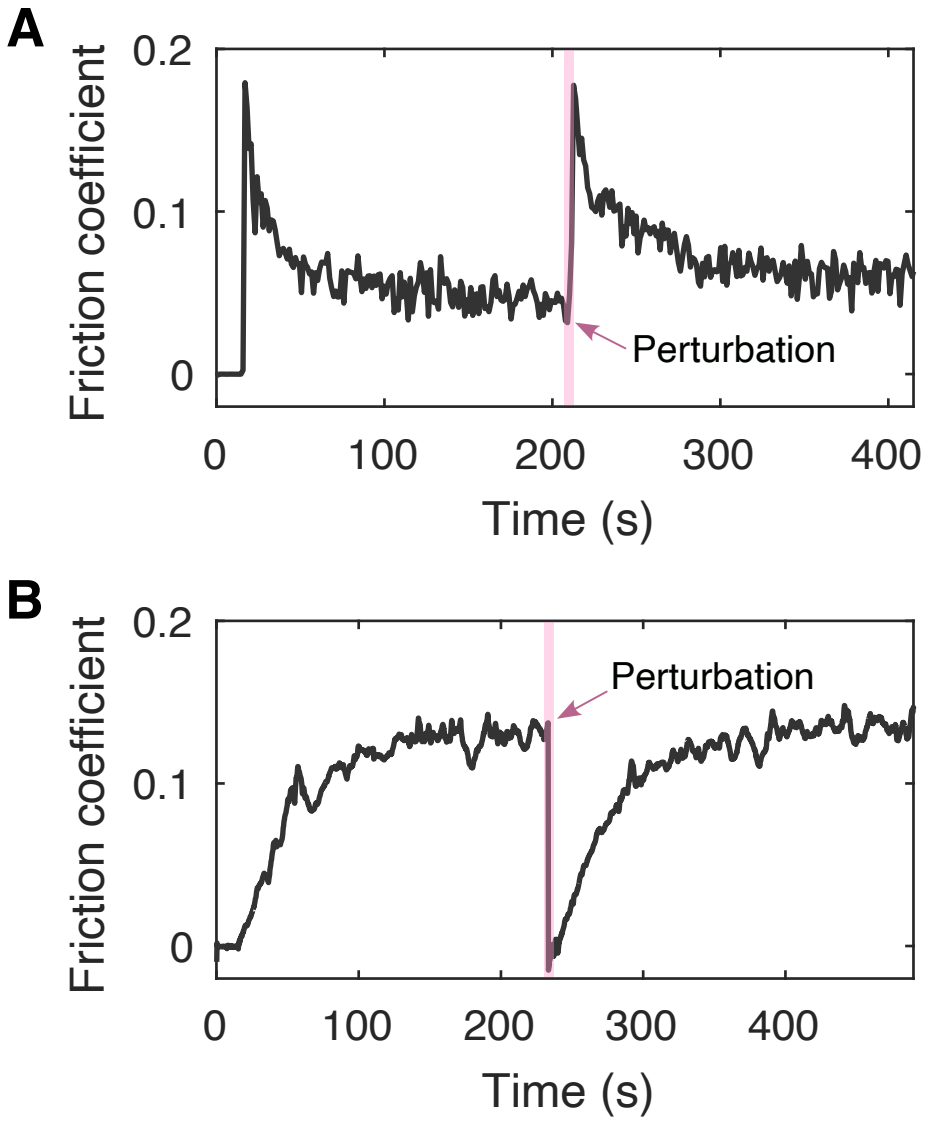

**Fig. 2. Mechanical perturbations regenerate frictional onset transients during sliding.** (A,B) Friction coefficient versus time at imposed sliding velocities of 1 μm $s^{-1}$ (A) and 4 nm $s^{-1}$ (B). The decaying (weakening) and rising (strengthening) responses, respectively, resemble the friction transients described by rate-and-state friction at high and low sliding rates (14). In (A) the perturbation is applied during steady-state sliding; in (B), it is applied near the peak of an overshoot that would require a longer sliding displacement to fully develop. In both cases, the transient is regenerated without a stationary hold, consistent with disruption of the shear-prepared interfacial configuration and its subsequent reorganization under continued shear.

Fluorescence images of the contact region provide qualitative evidence that the projected contact morphology changes with shear preparation (Fig. 3). For this purpose, the glass substrate is coated with fluorescence molecules as described elsewhere (25–27), which fluoresce more brightly when confined within a real contact; images were acquired in a separate setup with the rheometer mounted above a fluorescence microscope (25). Following separation and recontact, the interface displays a finely structured pattern containing many small, distinct contact patches (Fig. 3A). After steady sliding is established and subsequently arrested, the pattern appears less finely divided, with fewer resolved and more contiguous patches (Fig. 3B). Separating and recontacting the surfaces restores the more finely structured pattern (Fig. 3C). The projected contact morphology therefore correlates with the presence of the broad overshoot: the overshoot occurs upon restart from the finely structured states shown in Fig. 3A and Fig. 3C, but is absent from the less finely divided shear-prepared state shown in Fig. 3B.

These observations are consistent with our interpretation that the broad overshoot is associated with shear-induced reorganization of the asperity-contact population. The transition from many small, separated patches to fewer, more contiguous regions indicates that shear alters the projected distribution of load-bearing contacts across the interface. However, because the fluorescence images provide only a two-dimensional projection of a three-dimensional contact structure, they do not uniquely determine the microscopic spatial arrangement of individual asperity contacts. We therefore use these images as qualitative evidence for a preparation-dependent change in contact morphology, rather than as a quantitative reconstruction of the three-dimensional contact network. The recovery of the finely structured pattern after separation and recontact also argues against cumulative wear or irreversible flattening as the primary origin of the observed change. Creep and plastic deformation may nevertheless modify the asperity contacts over the experimental sequence and could account in part for the reduced overshoot amplitude after recontact (Fig. 1A,B).

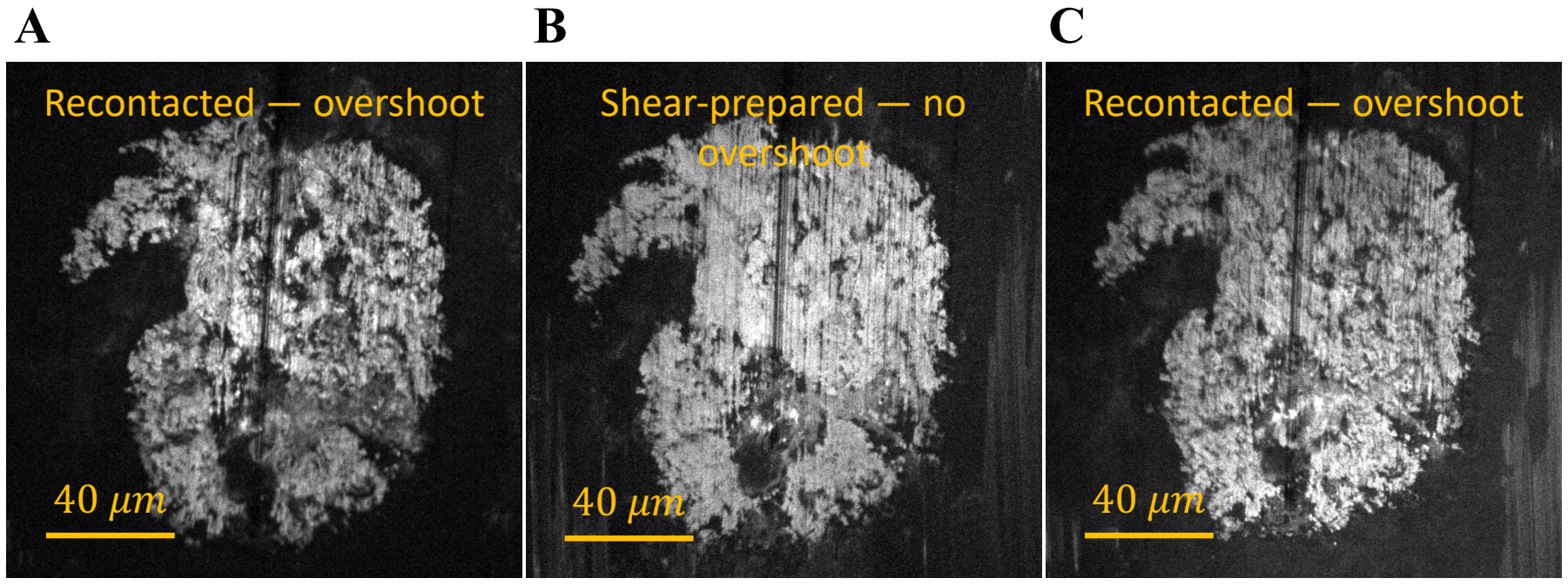

**Fig. 3. Projected contact morphology in three interfacial states.** Fluorescence images of a PTFE–glass contact, obtained using the fluorescence contact-imaging method (25–27), for three preparations within one sliding sequence. (A) Contact following separation and recontact, a preparation that produces the broad onset overshoot. (B) The same contact after steady sliding and subsequent arrest while retaining the shear-prepared state; sliding resumes from this preparation without the broad overshoot. (C) The contact after a further separation and recontact, which restores the overshoot. The shear-prepared state in (B) exhibits fewer resolved and more contiguous contact patches than the recontacted states in (A) and (C). The images are compared qualitatively, and projected contact area is not quantified. Recovery of the finely structured pattern after recontact is consistent with a reversible, preparation-dependent reorganization of the asperity-contact population.

## Distinguishing friction overshoot from contact aging

The same slide–pause–slide protocol can also reveal contact aging. Figure 4A shows a typical aging peak in the PTFE–glass system following a 20-min hold in a slide–pause–slide experiment, similar to Fig. 1A at $t = 10{,}000$ s, but with a longer pause. The peak height increases logarithmically with pause duration (Fig. 4B), as commonly observed in frictional aging (2, 6, 8, 11, 12). In compliant interfaces, this aging is commonly attributed to viscoelastic or plastic relaxation and creep of load-bearing asperities, often accompanied by an increase in the real area of contact (2, 5, 6, 8). The correspondence between contact area and frictional strength is not general, however: direct measurements show that the two can evolve independently, and that slip can increase the real contact area even as friction drops (10, 25). In our polymer interfaces specifically, we previously showed that frictional aging can be predicted from the bulk stress-relaxation response of the constituent material (8). In harder materials, contact aging has been associated with the gradual formation and strengthening of interfacial bonds at asperity microcontacts (16, 17).

These aging mechanisms strengthen load-bearing microcontacts that persist during the stationary hold (5, 6, 8, 16, 17). During sliding, by contrast, the population of microcontacts is continuously renewed as junctions deform, rupture, and reform. Direct observations of frictional interfaces have shown that slip replaces the population of microcontacts over a characteristic sliding distance (5). The broad overshoot studied here develops during this displacement-driven evolution. Moreover, viscoelastic creep or bond strengthening changes the size or strength of persistent junctions, but neither is expected to produce the shear-driven reorganization of the force-bearing network considered here — and the aging peak indeed appears without restoring the broad overshoot.

Unlike contact aging, the broad overshoot is restored when the shear-prepared configuration is disrupted, either by separating and recontacting the surfaces or by mechanical perturbation. Figure 4A also reveals the markedly different characteristic slip distances of the two responses: less than 200 nm for the aging peak, compared with roughly 10 μm for the broad overshoot. These observations demonstrate that the contact-aging peak and the broad overshoot are mechanically distinct responses associated with separable modes of interfacial evolution.

We previously interpreted the broad onset overshoot as arising from competition between contact aging and shear-induced rejuvenation (7), in line with the broader framework of such descriptions of frictional transients (2, 5, 11, 12, 19, 20). The experiments available at the time, however, could not separate these contributions or determine whether contact aging itself produced the broad overshoot. The present protocols show that this interpretation combined two distinct responses. Conventional contact aging produces the narrow, hold-time-dependent restart peak, whereas the broad overshoot is controlled by whether the shear-prepared interfacial configuration is preserved or disrupted. The broad transient remains well described by the expression introduced previously (7), which we recover from a phenomenological model combining relaxation toward steady sliding with a decaying, slip-dependent transient contribution (Text S1 and Fig. S4).

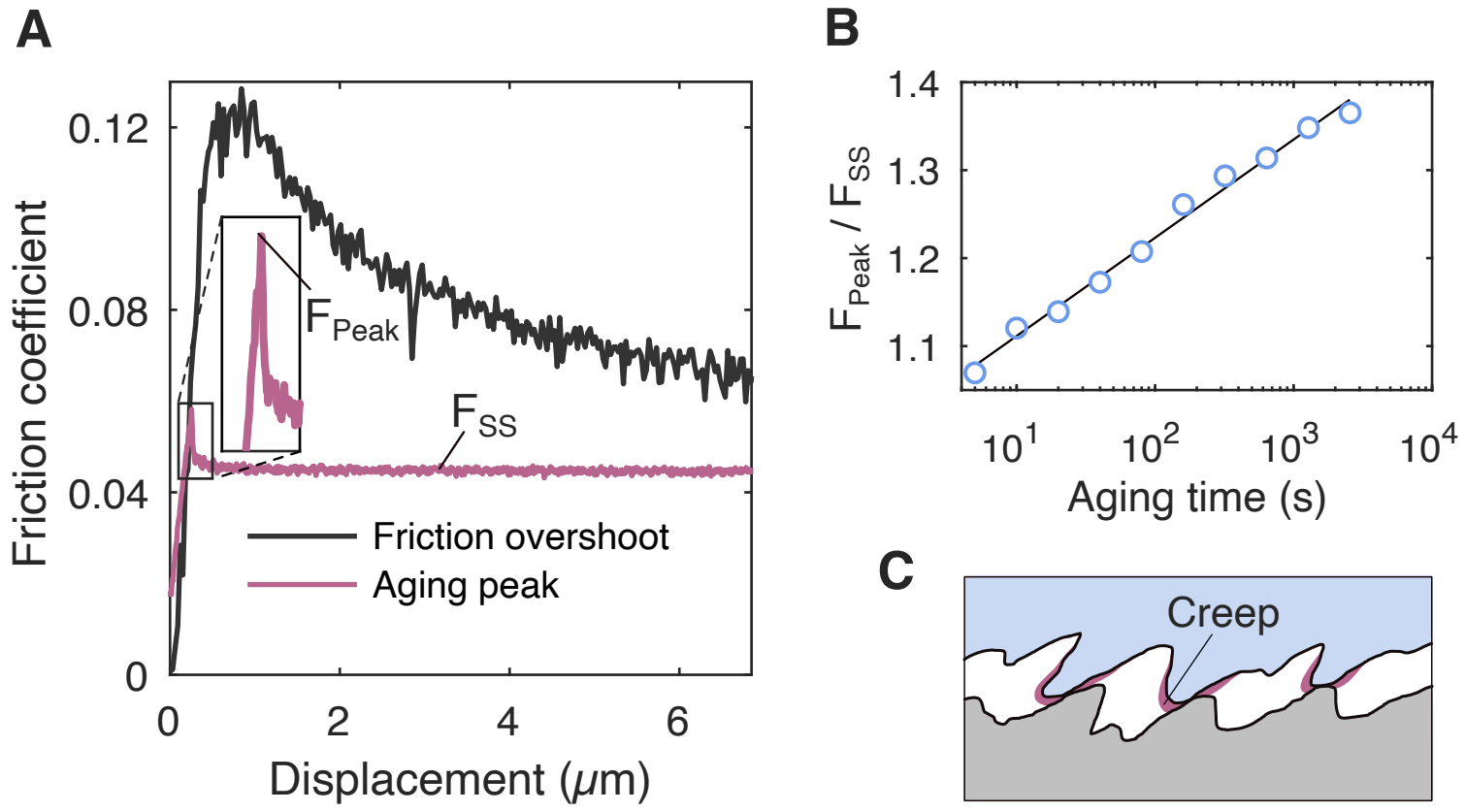

**Fig. 4. Friction overshoot versus contact aging.** (A) A typical aging peak from a slide–pause–slide experiment (20-minute pause) on the PTFE–glass interface, shown against displacement alongside the friction overshoot of Fig. 1A. The aging peak relaxes over roughly 200 nm, the overshoot over roughly 10 μm. (B) Normalized aging peak height ($F_{Peak}/F_{SS}$) versus pause duration; the solid line indicates logarithmic growth. (C) Schematic of viscoelastic creep in the softer asperities during the pause. Creep deforms individual asperities and increases the real contact area; it does not, however, relocate asperities within the interface.

## Discussion

Our experiments show that rough frictional interfaces evolve through at least two experimentally separable modes. During stationary contact, persistent microcontacts strengthen with elapsed time (2, 5, 6, 8, 16, 17). Under shear, the force-bearing contact network reorganizes with accumulated slip and shear direction (5, 7, 18). These processes can coexist, but they can be separated experimentally. Resetting or mechanically perturbing the interface regenerates the broad overshoot without an aging hold, whereas holding a shear-prepared interface at rest produces a narrow aging peak without restoring the broad overshoot. The two responses also relax over markedly different slip distances. The peak resistance at sliding onset can therefore reflect predominantly aging memory, or predominantly configurational memory, depending on how the interface was prepared.

Our findings also complement the shear-induced erasure of aging memory reported by Dillavou and Rubinstein (28). Their experiments showed that changes in static shear can partially erase the memory accumulated during a stationary hold. We find that sustained sliding prepares an additional configurational memory that survives arrest. The interface can subsequently continue to age while retaining the shear-prepared configuration that suppresses the broad overshoot. Shear can therefore both modify an existing aging memory and create a distinct configurational memory of its own. Taken together with their result, this suggests that shear acts on the interface in two ways — modifying the accumulated aging state, and preparing a configuration that constitutes a memory in its own right.

This distinction between aging and configurational memories has consequences for constitutive descriptions of friction. A one-state-variable rate-and-state law may remain an effective description over restricted classes of loading histories (11–13, 21, 22, 29). Multiple state variables are already permitted within the RSF framework (13), and two-state-variable formulations have been shown to outperform one-state descriptions in a low-stress, velocity-weakening regime (30). What our experiments provide is a physical basis for distinguishing such state information: temporal and configurational memories can be varied separately. A single scalar state variable should therefore not generally be identified with contact age, real contact area, or configurational organization alone. A physically interpretable description of the full set of protocols may instead require at least two state coordinates, or an equivalent higher-dimensional representation. One coordinate could describe the time-dependent strengthening of persistent contacts, whereas another could encode the evolution of the shear-prepared contact network with slip and shear direction.

The regeneration of the broad overshoot by a brief mechanical impulse further shows that frictional onset depends not only on hold time and shear history, but also on perturbation history. This sensitivity may be relevant to engineered and geophysical contacts exposed to ambient vibrations or transient mechanical disturbances (31, 32). More generally, the peak resistance commonly identified as static friction is not defined by a single interfacial state (9, 10). It is a protocol-dependent response of an interface that retains multiple forms of memory, a behavior recognized across a range of complex and disordered systems (3, 33, 34). Distinguishing the temporal memory written during stationary contact from the configurational memory written by shear provides a physical basis for interpreting frictional-onset measurements and for developing constitutive laws whose state variables correspond more directly to the microscopic processes they represent.

## Materials and Methods

### Friction apparatus

Friction experiments were conducted using an Anton Paar MCR 302 rheometer equipped with an electrically commutated (EC) motor that generates and measures torque in the nanonewton-meter range. This instrument enables exceptionally low rotational speeds down to $10^{-8}$ $s^{-1}$, equivalent to one full rotation every three years. A test sphere is mounted using a custom-built holder at radial distance $r = 7$ mm from the axis of rotation (Fig. S1). The rheometer presses the sphere against a glass substrate and drives it along a circular path while measuring the normal and friction forces. The rheometer is mounted on an optical vibration-isolation table (Newport) to minimize environmental vibrations. Such isolation is essential, as ambient vibrations can influence the configurational evolution of the asperities during measurements.

### Sample materials and surface preparation

Polypropylene spheres (2.45 mm diameter), polytetrafluoroethylene (PTFE) spheres (3.18 mm diameter), and steel ball bearings (2 mm diameter) were used. The PTFE and polypropylene spheres had root-mean-square surface roughness values of approximately 1.6 µm and 0.5 µm, respectively. The steel spheres were roughened by immersion in 20 wt% hydrochloric acid (HCl) for 15 min at room temperature, which etched the surface uniformly, yielding a roughness of approximately 0.4 µm. Surface roughness was measured using a 3D laser scanning microscope (Keyence VK-X1000).

Before each experiment, spheres were cleaned with ultrapure water (Milli-Q) and ethanol, then dried under nitrogen flow. Glass substrates were rinsed with Hellmanex cleaning solution, ultrapure water, and ethanol. This rinse cycle was repeated three times, followed by drying under nitrogen flow.

### Experimental protocols

*Data acquisition.* Friction measurements were recorded at 2–5 Hz sampling rate with 16-bit resolution. Raw data were used without filtering or averaging.

*Mechanical perturbations.* Mechanical perturbations were applied by dropping a 200-g sandbag (rubber balloon filled with fine sand) from 20 cm height onto the experimental table during steady-state sliding (Figs. 2A,B). This brief impulse perturbed the frictional interface through transmitted vibrations, without direct contact or changes to loading conditions.

*Elastic deformation correction.* The sample sphere and holder (Fig. S1) are not perfectly rigid and undergo elastic deformation under applied shear force. The rheometer measures the sum of interfacial sliding displacement and elastic deformation of the sphere and its holder. We extract the system's elastic compliance from the linear force–deformation behavior observed when sliding resumes in slide–pause–slide sequences (Fig. S4A). True interfacial sliding is obtained by subtracting this elastic contribution from the raw displacement data.

### Acknowledgments

We thank B. Weber, N. Ribe, and A. Davaille for helpful discussions. This work was funded by the European Research Council (ERC) under the European Union's Horizon 2020 research and innovation program (Grant Agreement No. 833240).

**Competing interests:** The authors declare no competing financial or non-financial interests.

SUPPLEMENTARY MATERIALS

**Text S1. Phenomenological description of the broad friction overshoot**

We describe the broad friction overshoot at sliding onset in terms of the accumulated interfacial sliding displacement $x$. At a fixed driving rate, we assume that continued sliding brings the interface toward a stable steady-state friction force $F_{\mathrm{SS}}$. In this asymptotic limit, the statistical properties of the interface become stationary and $dF/dx \to 0$. As a reduced description of the approach to steady sliding, we write

$$\frac{dF}{dx} = h(x) - g(F - F_{\mathrm{SS}}). \qquad \text{(S1)}$$

Here, $g$ represents a restoring contribution that depends on the deviation from steady-state friction, $\delta F = F - F_{\mathrm{SS}}$, while $h(x)$ represents a slip-dependent transient contribution. We take $g(0) = 0$ and assume that the restoring response is smooth and linearly stable near steady state. Expanding in $\delta F$ therefore gives

$$g(\delta F) = B\delta F + O(\delta F^2), \qquad B > 0. \qquad \text{(S2)}$$

The steady-sliding limit also requires $h(x) \to 0$ as $x \to \infty$. To specify its decay, we additionally assume an asymptotic expansion in inverse displacement with a nonzero leading term,

$$h(x) = \frac{A}{x} + O(x^{-2}), \qquad A > 0. \qquad \text{(S3)}$$

Retaining these leading terms gives

$$\frac{dF}{dx} = \frac{A}{x} - B(F - F_{\mathrm{SS}}), \qquad \text{(S4)}$$

which has the same mathematical form as the equation introduced in ref. (7). Here, $A$ has units of force, and $B^{-1}$ is a characteristic slip length associated with the restoring term. We use Eq. (S4) as a phenomenological description of the resolved broad transient, without identifying $A/x$ with strengthening during stationary contact.

The solution can be written as

$$\mathrm{F}(x) = Ae^{-Bx} \int_{x_0}^{x} \frac{e^{Bx'}}{x'} dx' + F_{SS}(1 - e^{-Bx}), \qquad \text{(S5)}$$

The parameter $x_0 > 0$ is the integration constant and represents the initial condition within this reduced description. Within this model, the transient contribution can raise friction above $F_{\mathrm{SS}}$, while the restoring term governs its subsequent decrease. At an extremum point $x_p$,

$$F(x_p) - F_{\mathrm{SS}} = \frac{A}{Bx_p} > 0. \qquad \text{(S6)}$$

Differentiating Eq. (S4) gives

$$\frac{d^2F}{dx^2} = -\frac{A}{x^2} - B\frac{dF}{dx},$$

so that

$$\frac{d^2F}{dx^2}|_{x_p} = -\frac{A}{x_p^2} < 0.$$

Thus, any extremum point of the model response is a maximum above the steady-state friction.

To compare Eq. (S5) with experiment, we correct the measured displacement for elastic deformation of the sample sphere and its holder. The rheometer displacement contains both interfacial slip and elastic deformation under shear. We determine the effective shear stiffness $k$ from the initial linear force–displacement response observed when sliding resumes after a brief pause. The true interfacial slip is then obtained as $x = x_{\text{meas}} - F/k$. The correction procedure is illustrated in Fig. S4A.

Fits to the corrected data are shown in Figs. S4B–D. The expression captures the initial increase and subsequent relaxation of the broad transient over the measured displacement range. The initial and regenerated PTFE overshoots are described using common values of $\mu_k = F_{\text{SS}}/N$, $B$, and $A/N$, with different values of $x_0$. Within the model, their preparation dependence is therefore represented through the initial condition rather than changes in the shared evolution parameters.

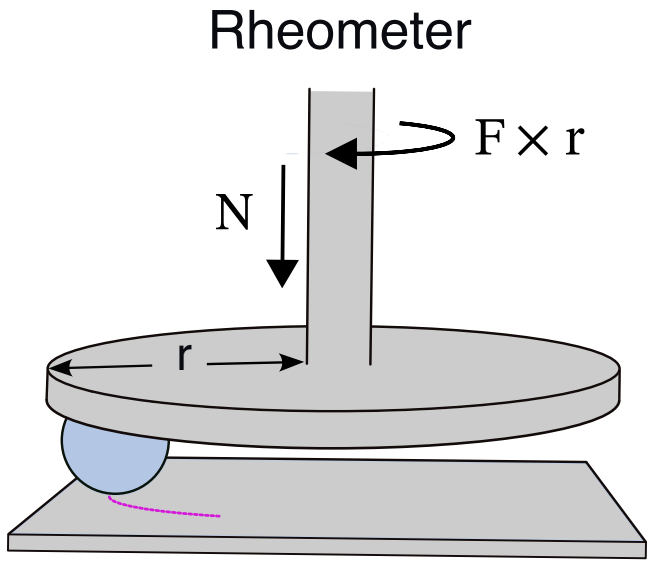

**Fig. S1. Schematic of the experimental setup.** A sample sphere is mounted on a steel disk via a holder at radial distance $r = 7$ mm from the disk center. The disk is connected to a rheometer that presses the sphere against a glass substrate with normal force $N$ and drives it along a circular path by applying torque $F \times r$, where $F$ is the friction force.

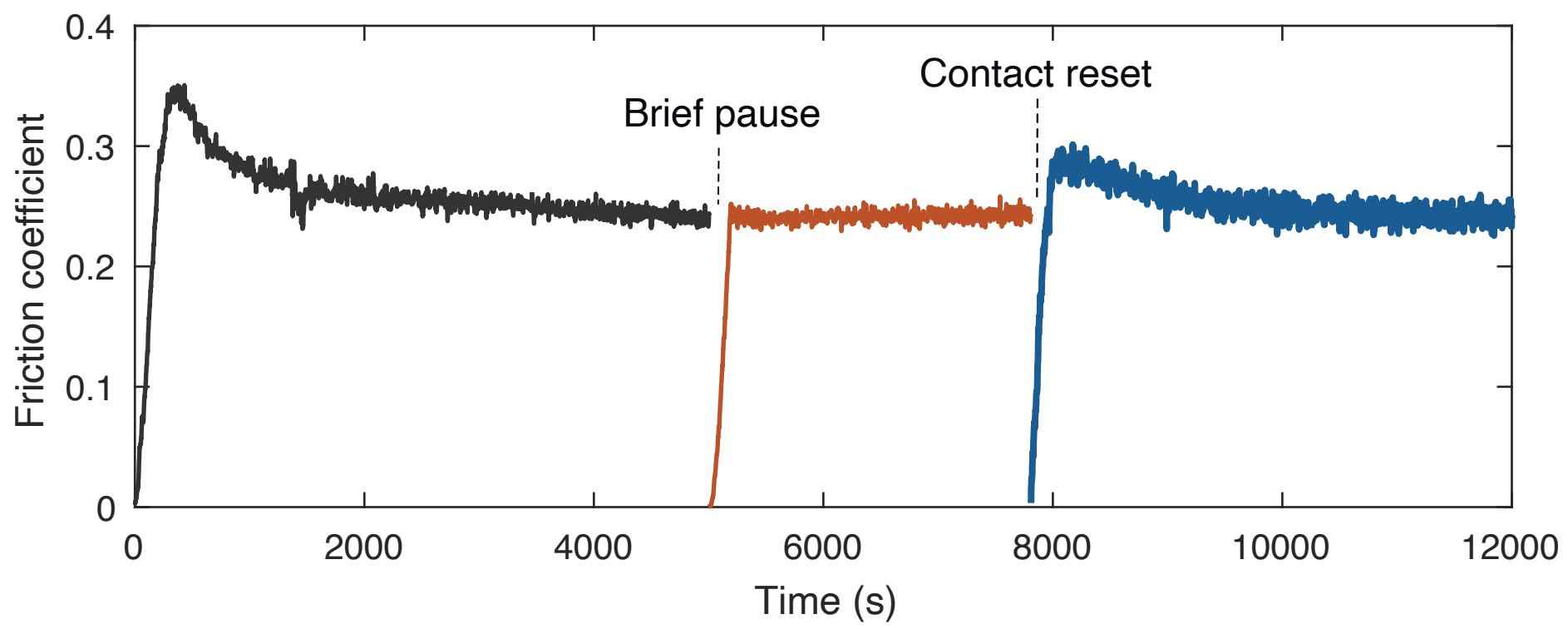

**Fig. S2. Configurational dependence of the friction overshoot at a polypropylene–glass interface.** Friction coefficient versus time for a polypropylene–glass interface under an imposed sliding velocity of 4 nm $s^{-1}$. After reaching steady-state sliding, sliding was paused for 5 s and then resumed. At a later time, sliding was stopped, surfaces were separated and re-contacted, and sliding restarted. The brief pause does not restore the broad overshoot, whereas separation and recontact does.

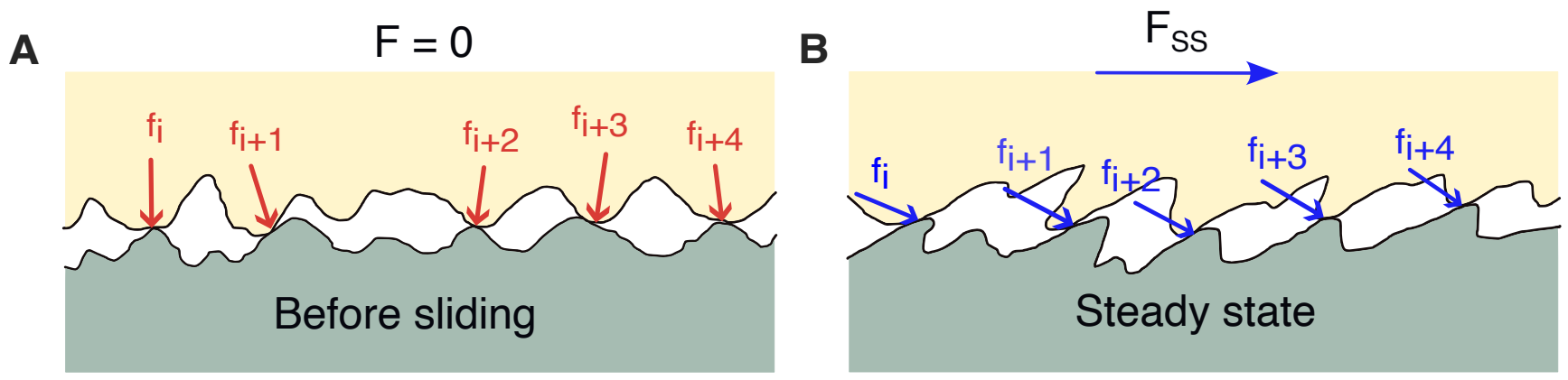

**Fig. S3. Proposed configurational picture of frictional onset.** Schematic of asperities at a rough interface before sliding (left) and during steady-state sliding (right). The macroscopic friction force along the sliding direction is $F = \sum_i f_{x,i}$, where $f_{x,i}$ is the sliding-direction component of the force transmitted by asperity $i$. In this picture, initially disordered asperity forces produce no net force along the sliding direction; under shear, these forces' contributions become progressively biased along the sliding direction, generating macroscopic kinetic friction at steady state.

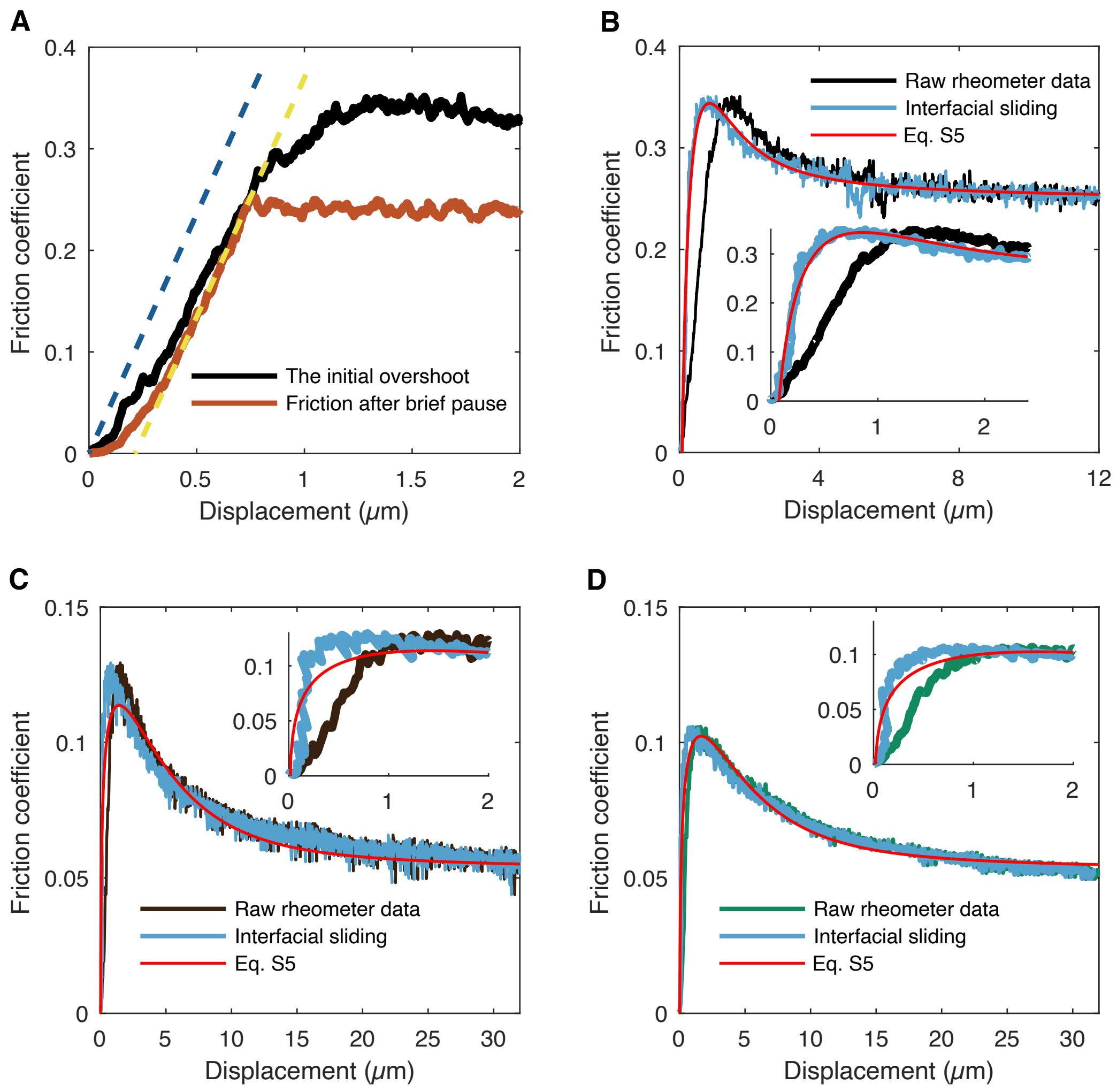

**Fig. S4. Empirical description of the friction overshoot.** (A) Polypropylene–glass friction data from Fig. S2 plotted against the total rheometer displacement during the initial overshoot (black) and after a brief pause (brown). The yellow dashed line follows the measured linear elastic response after the pause, and the blue dashed line shows the same response translated to the origin; the horizontal distance from the blue line to the measured curve gives the interfacial slip at each force level. The effective shear stiffness $k$ of the sphere and holder is determined from this response, and the elastic contribution is subtracted using $x = x_{meas} - \frac{F}{k}$ to obtain the interfacial slip. (B) Polypropylene–glass data from Fig. S2, showing the raw data plotted against measured rheometer displacement (black) and the interfacial-slip curve after elastic correction (blue). The corrected data are fitted with Eq. (S5): $\mu_k = F_{SS}/N = 0.25$, $B = 1.93$ μm$^{-1}$, $A/N = 0.16$, and $x_0 = 0.094$ μm. (C,D) PTFE–glass data for the initial overshoot (C) and the regenerated overshoot after contact reset (D) from Fig. 1A. The elastically corrected interfacial-sliding curves are shown in blue and fitted with Eq. (S5). The fitted parameters are $\mu_k = 0.052$, $B = 0.32$ μm$^{-1}$, $A/N = 0.027$, with $x_0 = 0.011$ μm in (C) and $x_0 = 0.016$ μm in (D).